\documentclass[conference]{IEEEtran}
\usepackage{cite}
\usepackage{amsmath,amssymb,amsfonts}
\usepackage{algorithmic}
\usepackage{graphicx}
\usepackage{textcomp}
\usepackage{xcolor}
\usepackage{url}
\usepackage{listings}
\usepackage{booktabs}
\usepackage[T1]{fontenc}

\graphicspath{{figures/}}

\begin{document}

\title{WVA: A Global Optimization Control Plane for \texttt{llmd}}

\author{
    \IEEEauthorblockN{
        Abhishek Malvankar\IEEEauthorrefmark{1},
        Lionel Villard\IEEEauthorrefmark{1},
        Mohammed Abdi\IEEEauthorrefmark{1},
        Tommaso Sgreccia\IEEEauthorrefmark{2},
        Evgeny Shindin\IEEEauthorrefmark{1},
        Braulio Dumba\IEEEauthorrefmark{1}, \\
        Vishakha Ramani\IEEEauthorrefmark{1},
        Asser Tantawi\IEEEauthorrefmark{1},
        Tamar Eilam\IEEEauthorrefmark{1}
    }
    \IEEEauthorblockA{\IEEEauthorrefmark{1}\textit{IBM Research} \\
    \{asmalvan, villard, tantawi, eilamt\}@us.ibm.com, \\
    \{mohammed.munir.abdi, braulio.dumba, vishakharamani\}@ibm.com, \\
    evgensh@il.ibm.com}
    \IEEEauthorblockA{\IEEEauthorrefmark{2}\textit{University of Bologna} \\
    tommaso.sgreccia@studio.unibo.it}
}

\maketitle
\thispagestyle{empty}
\pagestyle{plain}

\begin{abstract}
As Large Language Models (LLMs) scale to handle massive concurrent traffic, optimizing the infrastructure required for inference has become a primary challenge. To manage the high cost of GPU resources while ensuring strict service-level objectives (SLOs), operators increasingly deploy models across heterogeneous hardware clusters that multiplex latency-sensitive online requests and throughput-oriented offline requests. However, traditional resource-centric autoscalers like the Kubernetes horizontal pod autoscaler (HPA) do not consider application-specific SLOs, hardware heterogeneity, or internal engine state (like KV cache utilization) globally. This leads to unnecessary scaling, severe resource underutilization, and disrupted stateful inference. To address these limitations, we introduce the Workload Variant Autoscaler (WVA), a specialized control plane co-designed with \texttt{llmd} that tightly couples scaling decisions with the inference server's internal saturation state. By utilizing proactive headroom-based scaling and fragmentation-aware scale-down, our experiments demonstrate that WVA achieves a \textbf{37\% improvement in effective throughput} and a \textbf{10x reduction in request failures} compared to HPA. Furthermore, WVA's cost-aware tiering intrinsically reduces overall power consumption by prioritizing lower-cost, energy-efficient hardware variants over homogeneous scaling on high-end accelerators.
\end{abstract}


\section{Introduction}
The inference cost of serving Large Language Models (LLMs) has emerged as a critical bottleneck for the AI industry \cite{vllm, orca}. In contrast to stateless microservices where throughput scales linearly with compute resources, LLM inference is a memory-bound, stateful process. Performance is constrained by high-bandwidth memory (HBM) capacity for Key-Value (KV) caches and varies significantly based on input context length and decoding parameters \cite{vllm}. Legacy autoscaling mechanisms like the Kubernetes HPA, designed for the homogeneous clusters of the microservices era, are ill-equipped for this new reality. Generic autoscalers treat the application as a black box, optimizing for \textit{Resource Utilization} (e.g., keep CPU at 80\%). We argue that this abstraction is leaky for LLMs. High-performance inference requires white-box autoscaling, where the scaler deeply understands the engine's internal state.
This impedance mismatch leads to massive inefficiency. GPU clusters are often over-provisioned to handle peak load, wasting energy during idle periods, or they fail to distinguish between expensive H100s and cost-effective A100s, treating different types of hardware as fungible units. To solve this, we must shift from resource-based scaling to SLO-based, heterogeneity-aware scaling.

\begin{figure}[!t]
  \centering
  \includegraphics[width=0.85\columnwidth]{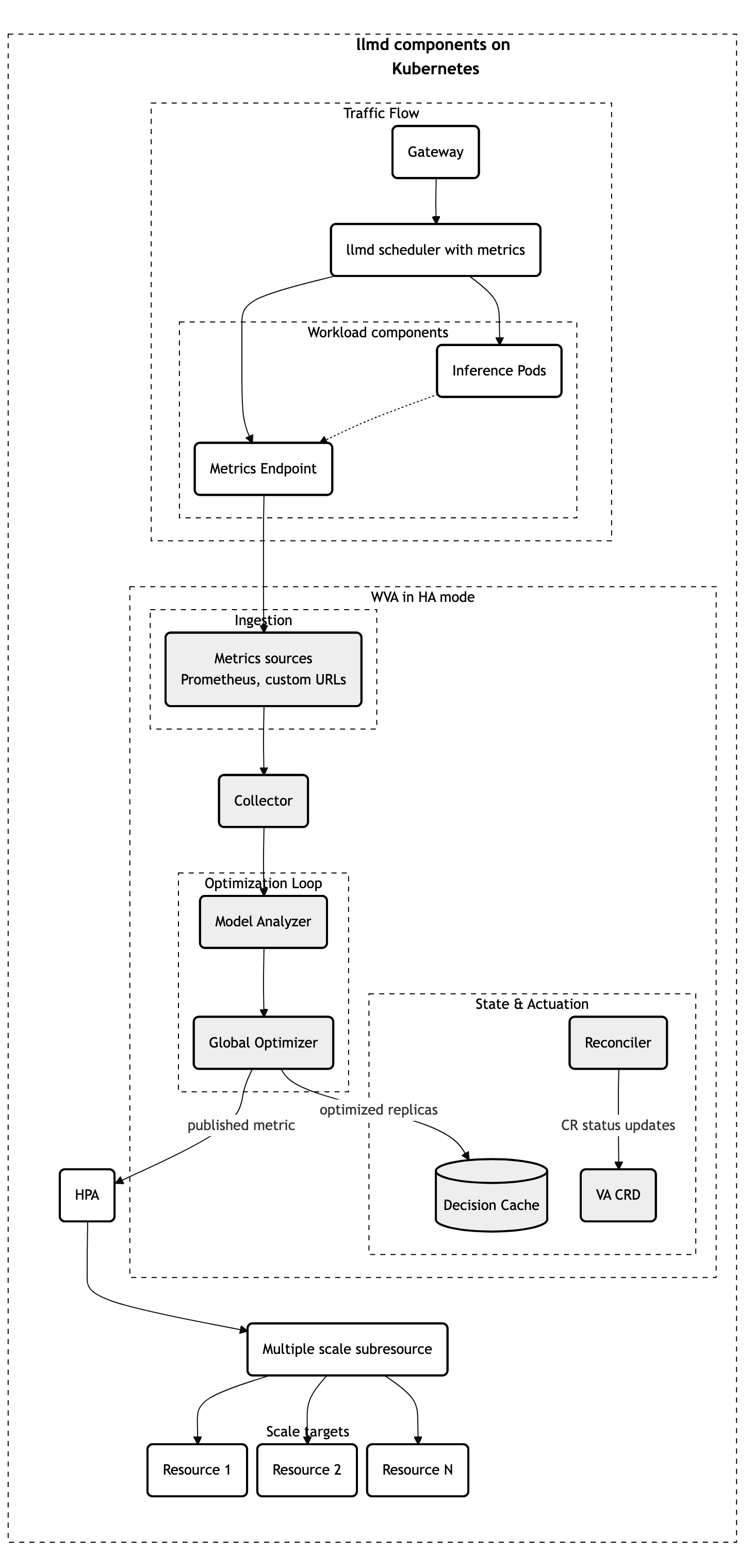}
  \caption{WVA integrated with \texttt{llmd}}
  \label{fig:design}
\end{figure}

We introduce the Workload Variant Autoscaler (WVA), a control plane framework designed for these complexities (see Figure \ref{fig:design}). WVA integrates deeply with \texttt{llmd}~\cite{llmd}, a Kubernetes-native high-performance distributed LLM inference framework, to improve latency and throughput while providing cost and energy efficiency. Unlike monolithic scalers, WVA features a pluggable architecture that allows operators to inject custom optimization strategies for conflicting objectives. Leveraging this framework, we implement a saturation-based optimization that enables advanced capabilities such as headroom-based scaling and fragmentation-aware scale-down.

\section{Background and Motivation}

\subsection{Background: LLM Inference Dynamics}
LLM inference is fundamentally distinct from traditional stateless microservices, characterized by its stateful nature and heterogeneous hardware requirements. The process comprises two distinct phases: \textit{Prefill} and \textit{Decode}. The prefill phase processes the input prompt in parallel to compute the initial Key-Value (KV) cache, a highly compute-bound operation. The decode phase subsequently generates tokens sequentially; this phase is memory-bandwidth bound and relies heavily on the cached KV states \cite{splitwise}. The \textit{KV Cache} stores these intermediate attention states to avoid redundant computation, but it consumes significant GPU memory, making memory capacity the primary bottleneck. 

This architectural complexity introduces significant variability in latency expectations. User requests vary drastically in input prompt length and desired output tokens, leading to unpredictable processing times. Furthermore, due to the extreme capital expenditure required for AI infrastructure, clusters are increasingly heterogeneous, mixing older, cost-efficient GPUs (e.g., NVIDIA A100s) with newer, highly performant but expensive GPUs (e.g., NVIDIA H100s).

\subsection{Motivation: Balancing Cost and SLOs}
The core motivation for our work is the urgent need to balance inference cost with stringent Service Level Objectives (SLOs). In production Model-as-a-Service (MaaS) environments, providers must navigate conflicting constraints: maximizing hardware utilization to reduce the per-token serving cost, while simultaneously maintaining strict tail-latency guarantees for end users. Achieving this balance requires an autoscaling control plane that can intelligently scale traffic across heterogeneous hardware tiers. For example, a system should prioritize scaling inexpensive A100 instances for baseline traffic to minimize cost, while reserving expensive H100 capacity to absorb sudden load spikes and protect latency SLOs. 

\subsection{State of the Art: The Shortcomings of Legacy Scaling}
Legacy autoscaling systems, such as the Kubernetes horizontal pod autoscaler (HPA), are insufficient for the complexities of LLM inference because they were designed for homogeneous, stateless workloads. Simple HPA operates on aggregated, generic resource metrics (e.g., CPU or memory utilization) and optimizes for a target average \textit{Resource Utilization} (e.g., maintaining CPU at 80\%). This approach treats the inference engine as a black box and fails to capture application-specific congestion signals like KV cache fragmentation or active queue depth.

Even if HPA is extended via custom metric adapters, it still fundamentally treats all pods as identical, fungible replicas. It offers no native mechanism for cost-aware tiering (e.g., prioritizing A100s over H100s). Finally, HPA's scale-down logic results in agnostic termination, removing replicas based on cluster-wide averages without regard for locally saturated KV caches on specific nodes, which severely disrupts stateful inference. Ultimately, solving these challenges requires the deep vertical integration of application-aware saturation, heterogeneity support, and decoupled actuation that legacy systems cannot provide.

A common theoretical alternative is to modify the core Kubernetes HPA to natively understand these LLM-specific paradigms. However, doing so actively contradicts the cloud-native philosophy of modularity. HPA is intentionally designed as a generic, lowest-common-denominator controller for stateless, fungible pods. Embedding domain-specific logic---such as KV cache phase awareness, prefill/decode scaling requirements, or heterogeneous hardware tiering---into the core control plane would break backward compatibility and severely bloat the orchestrator. As highlighted by emerging CNCF initiatives \cite{llmd_sandbox}, managing the unique resource-utilization asymmetry of generative AI requires specialized, decoupled data-plane extensions rather than monolithic alterations to foundational orchestration primitives. Consequently, WVA is designed as an extensible, inference-aware overlaid control plane rather than a fork of HPA.
\section{System Model}

\subsection{The Variant Abstraction}
Central to WVA is the introduction of the Variant as a first-class abstraction. A single base model (e.g., `Llama3-70b`) could be deployed in multiple inference server configurations. We define a \textit{Variant} as a unique tuple of:
\begin{equation}
    Variant = \langle Hardware, Parallelism, Quantization \rangle
\end{equation}
For the current iteration of the WVA implementation evaluated in this work, we specifically constrain the variant space to two primary dimensions: \textit{Hardware}, denoting the specific GPU model architecture hosting the inference server, and \textit{Parallelism}, representing the discrete number of GPUs allocated per model replica. For example, `Variant A` might be `(H100, 2 GPUs)` while `Variant B` is `(A100, 4 GPUs)`. This abstraction is operationalized via the \texttt{VariantAutoscaling} (VA) control-plane resource, enabling the system to evaluate the \textit{cost} and \textit{performance} trade-offs of each variant independently, rather than treating the service as a homogeneous pool of pods.

\subsection{Saturation Model (Default Strategy)}
WVA functions as the autoscaling companion to \texttt{llmd}, consuming granular metrics such as \texttt{queue\_length} and \texttt{kv\_cache\_usage}. Unlike generic resource metrics (e.g., CPU utilization), these signals provide a direct measure of application-level congestion, allowing the autoscaler to differentiate between compute-bound processing and memory-bound capacity limits.

Although WVA supports diverse strategies through the \texttt{Auto\allowbreak scaling\allowbreak Policy} interface, we formulate our core saturation-based optimization as a safety-margin reactive control loop. Let $R$ denote the set of all active service replicas. For each measurable resource metric $m \in \{kv, q\}$ (representing KV cache usage and queue depth, respectively), we define a hard saturation threshold $\tau_{m}$. This threshold $\tau_{m}$ represents the critical utilization boundary—derived via offline profiling of the specific model variant—beyond which request latency degrades non-linearly.

We formally define the subset of saturated replicas $S \subseteq R$ as those whose instantaneous utilization $U_{m}(r)$ for any metric $m$ meets or exceeds its corresponding threshold $\tau_{m}$:
\begin{equation}
    S = \{ r \in R \mid U_{kv}(r) \ge \tau_{kv} \lor U_{q}(r) \ge \tau_{q} \}
\end{equation}
where $U_{kv}(r)$ and $U_{q}(r)$ denote the instantaneous KV cache usage and queue depth of replica $r$.
The engine calculates the average spare capacity $\delta_{avg}$ across the non-saturated set $R \setminus S$. A scale-up event is triggered if this safety margin drops below the configured trigger $\gamma$:
\begin{equation}
    \exists m \in \{kv, q\} : \frac{1}{|R \setminus S|} \sum_{r \notin S} (\tau_{m} - U_{m}(r)) < \gamma_{m} \label{eq:trigger}
\end{equation}
Conversely, scale-down is only permitted if the projected spare capacity after removing a replica remains above $\gamma$, ensuring stability.

Role Awareness: \texttt{llmd} distinguishes between nodes handling prefill (compute-bound) vs. decode (memory-bound). WVA's \texttt{AutoscalingPolicy} matches these roles, applying different scaling aggression profiles.

\subsection{Global Optimization Framework}
WVA's Control Plane Framework allows for global optimization strategies unattainable by local, pod-centric autoscalers.

\subsubsection{Headroom-Based Capacity Management} \label{sec:headroom}
We define Headroom ($\delta$) as the aggregate spare capacity available within the cluster to process new requests immediately. While Equation~\ref{eq:trigger} from Section III.B utilizes the threshold $\gamma$ to trigger an abstract scaling need, we formalize $\delta_{safety}$ as the concrete, quantifiable volume of cluster-level capacity required to satisfy that safety margin. Unlike HPA, which reacts to deviations from a target utilization (e.g., scaling only after load exceeds 80\%), WVA formulates capacity provisioning as an optimization problem. Let $Load_{current}$ denote the aggregate resources globally consumed by active traffic. We formally define the optimal system state, $Capacity_{target}$, as the absolute resource allocation required to process $Load_{current}$ while strictly maintaining the reserved safety buffer $\delta_{safety}$:
\begin{equation}
    Capacity_{target} = Load_{current} + \delta_{safety} \label{eq:target}
\end{equation}
The Global Optimizer ensures that this spare capacity buffer exists across the cluster to absorb instantaneous traffic bursts without incurring provisioning latency. Formally, Equation~\ref{eq:trigger} and Equation~\ref{eq:target} constitute a two-stage reactive-proactive control loop for headroom management. Equation~\ref{eq:trigger} defines the state-dependent trigger condition, continuously evaluating the instantaneous spare capacity against a critical lower bound. Upon violation of this condition, the system transitions to the actuation phase, wherein Equation~\ref{eq:target} computes the requisite target allocation to globally restore the safety margin $\delta_{safety}$. This decoupled formulation allows WVA to mirror the routing heuristics of the inference scheduler (EPP) inside \texttt{llmd}, guaranteeing that scaling decisions are phase-aligned with request admission, thereby mitigating control loop oscillation.

\subsubsection{Fragmentation-Aware Safety}
In heterogeneous distributed systems, load is rarely uniformly distributed. Purely reactive scalers suffer from distributional masking, and if the cluster-wide average utilization is low, they may terminate replicas. However, in stateful inference, specific pods may still be saturated (fragmented usage) based on scorer(s) enabled in \texttt{llmd} scheduler. WVA's global optimizer aggregates per-replica saturation signals to construct a precise view of cluster state. It permits scale-down operations only when specific variants are inextricably identifiable as idle. To further mitigate the risk of cascading saturation during load redistribution, the optimizer enforces a strict lower bound (\texttt{MinNonSaturatedReplicasForScaleDown}, defaulting to 2) on the number of globally non-saturated replicas that must remain active before any scale-down event is authorized. This combined approach serves as a deterministic safety mechanism against data loss or interruption of long-running requests.

\subsection{End-to-End Workflow}
While we omit the detailed request processing workflow of \texttt{llmd}, typically requests are routed to a Gateway, which forwards them to an \texttt{llmd} scheduler to select the least saturated pod. Once requests reach the inference server pod, WVA performs the following steps to maintain optimal capacity (see Figure~\ref{fig:design}):
\begin{enumerate}
    \item Metric Ingestion: The optimization cycle (model analyzer and global optimizer) consumes high-fidelity stats (KV cache usage, queue depth) from inference servers via the Collector.
    \item Optimization Loop: The engine calculates the optimal distribution of replicas across variants (e.g., A100 vs. H100) using the Model Analyzer with optional global optimization in Constrained Mode.
    \item Decision Caching: The target state is written to the in-memory \texttt{DecisionCache}.
    \item Reconciliation: Triggered by the decision engine or VA CRD updates, the Reconciler reads the cache and patches the \texttt{VariantAutoscaling} Status with the new optimized allocation.
    \item Actuation: The controller updates the underlying Kubernetes objects that has scale subresource to match the target replica count via HPA.
\end{enumerate}

\subsection{Configuration Interface}
Users instantiate optimization for specific workloads by declaring \texttt{VariantAutoscaling} (VA) Custom Resources. Each VA resource binds a logical model variant to a physical Kubernetes target. The core specification fields include:
\begin{description}
    \item[Model Identifier] \hfill \\
    A unique ID (e.g., \texttt{llama3-70b}) specified in the \texttt{modelID} field, used to group multiple VA resources into a single pool for capacity planning.
    \item[Variant Cost] \hfill \\
    A scalar string value (field \texttt{variantCost}, default "10.0") representing the relative cost per replica. The solver saturates lower-cost variants first.
    \item[Scale Target] \hfill \\
    The target Deployment, referenced via the \texttt{scaleTargetRef} field.
\end{description}

Saturation thresholds (e.g., queue depth limits) are managed centrally via a configuration map, allowing operators to tune sensitivity without redeploying individual variant resources.

\section{Architecture}

\subsection{Architectural Overview}
WVA is built to optimize the latency and throughput of the \texttt{llmd} inference platform. A key requirement is supporting numerous base models that must scale to zero during idle periods but scale up immediately upon traffic arrival without request loss. Another requirement was to scrape metrics natively from vLLM inference servers and the \texttt{llmd} scheduler (EPP - endpoint picker) pod rather than relying on Kubernetes machinery to publish events. WVA is designed as a flexible control plane framework that accelerates the development of new scaling strategies. To achieve these goals, WVA comprises four primary components:
\begin{description}
    \item[Reconciler] \hfill \\
    Responsible for processing variants and updating their state, including managing the scale-to-zero lifecycle.
    \item[ScaleFromZero] \hfill \\
    A dedicated thread that operates at a higher frequency than the standard Reconciler loop. This component addresses the MaaS requirement for latency-critical cold starts, ensuring immediate responsiveness to initial traffic to prevent request drops.
    \item[Collectors] \hfill \\
    A modular framework that allows for the registration of different collector implementations and query types. This modularity enables WVA to flexibly ingest metrics from diverse backends (e.g., Prometheus, local metrics endpoints) to drive decisions.
    \item[Decision Cache] \hfill \\
    A high-performance in-memory state store for optimization decisions. It decouples the optimization logic from the Kubernetes API server, ensuring updates are pushed only when necessary to maintain system stability.
    \item[Optimizers] \hfill \\
    A composite engine designed to run pluggable strategies. In our reference implementation, we deploy a saturation-based optimizer consisting of a model analyzer and global optimizer. These components work in tandem to optimize workload scaling for cost and energy efficiency.
\end{description}
This separation ensures that the critical path---scaling the inference server---is never blocked by slow metric backends.

To ensure seamless integration with standard Kubernetes operational workflows (e.g., GitOps, Helm), WVA's control plane is built on industry-standard foundations. We employ \texttt{viper} for hierarchical configuration management, enabling operators to dynamically override settings via flags, environment variables, or configuration with strict precedence logic. This design ensures that WVA can be easily embedded into diverse platform stacks without requiring bespoke configuration adaptors.

\subsection{Design Rationale: Why Pluggability?}
The decision to architect WVA as a modular framework is driven by three characteristics of the current AI landscape:

\begin{enumerate}
    \item Rapid Engine Evolution: Inference servers like vLLM and SGLang are changing continuously, introducing new features such as chunked prefill and speculative decoding that fundamentally alter bottlenecks. A rigid scaler scaling on distinct queue depth logic becomes obsolete with every engine upgrade. WVA's pluggable collectors allow operators to seamlessly shift scaling signals (e.g., from queue length to inter-token latency) as engines evolve.
    \item Divergent Workload Objectives: Online inference and offline batch workloads possess diametrically opposing scaling objectives. Latency-critical services require strict SLO adherence (scaling up \textit{before} saturation), while throughput-oriented offline tasks prioritize cost-efficiency (packing queues to maximize throughput). WVA's \texttt{AutoscalingPolicy} interface allows distinct optimization algorithms to be swapped per deployment, enabling a single control plane to manage both latency-critical and throughput-oriented clusters simultaneously.
    \item Hardware Diversity: The definition of saturation varies by hardware. A memory-bound model on an A100 requires KV-cache-based scaling, while the same model on a compute-dense H100 might be bound by tensor parallelism overhead~\cite{megatron}. Pluggable Optimizers enable hardware-specific scaling heuristics to be injected without modifying the core reconciliation loop.
\end{enumerate}

\subsection{Pluggable Interfaces}
To ensure WVA can adapt to diverse operating environments, we define clear Go interfaces for its core logic. This allows developers to inject custom scaling algorithms or metric backends without forking the core codebase.

\begin{lstlisting}
// AutoscalingPolicy calculates target state
// independent of how that state is applied.
type AutoscalingPolicy interface {
    // Returns calculated target replicas
    Calculate(signals SystemState) 
        (TargetState, error)
}

// ActuationStrategy supports different
// actuation mechanisms (e.g., HPA, APIs).
type ActuationStrategy interface {
    // Publishes decision for the orchestrator
    Publish(decision TargetState) error
}
\end{lstlisting}

This separation of concerns mirrors the design of extensible systems like Mesos~\cite{mesos}, where resource allocation is decoupled from task execution.

\subsection{Modular Metrics Collection}
Complementing the pluggable optimization logic, WVA employs a modular architecture for metric ingestion (the V2 Collector API), offering adaptable observability. This allows the system to be deployed in diverse environments---from air-gapped on-premise clusters to public cloud VPCs---without modifying the core autoscaler.

The system abstracts metric retrieval via the \texttt{Metrics\allowbreak Source} interface and a dynamic \texttt{Registry}.

\begin{lstlisting}
// MetricsSource abstracts retrieval
type MetricsSource interface {
    // Refresh fetches latest values
    Refresh(ctx context.Context, 
        spec RefreshSpec)
        (map[string]*MetricResult, error)

    // Get retrieves a cached value
    Get(queryName string, 
        params map[string]string) *CachedValue
}
\end{lstlisting}

Key architectural components include:
\begin{itemize}
    \item Pluggable Source Interface: A unified interface abstracts the data retrieval mechanism, allowing WVA to pull metrics from diverse sources (e.g., Prometheus, direct push APIs) without modifying the core engine.
    \item Registry Pattern: A central \texttt{Registry} dynamically instantiates configured metric sources at startup. This modularity not only simplifies testing via mock providers (e.g., \texttt{FileSource} for simulation) but also future-proofs the system against changes in the monitoring infrastructure.
    \item Abstraction Layer: The optimization engine operates on a normalized internal data model, completely decoupled from the specific query languages or storage formats of the underlying observability backends.
\end{itemize}

\subsection{Pluggable Resource Discovery}
Similar to the metric collection layer, the resource discovery layer is also decoupled behind interfaces, providing environment-agnostic discovery. This further reinforces the extensible framework narrative of WVA, as it does not assume a specific cluster topology or management layer.

\begin{lstlisting}
// CapacityDiscovery abstracts hardware inventory
type CapacityDiscovery interface {
    // Returns map of node->model->info
    Discover(ctx context.Context)
       (map[string]map[string]ModelInfo, error)
}
\end{lstlisting}

The default implementation, \texttt{K8s\allowbreak With\allowbreak Gpu\allowbreak Operator}, queries Kubernetes nodes for GPU Operator labels. However, this interface allows WVA to support non-Kubernetes environments (e.g., by reading from a static inventory file) or custom GPU management systems without modifying the core optimization logic.

\subsection{The Model Analyzer}
The core logic for local saturation detection resides in the Model Analyzer. It corresponds to the Model Analyzer component in Figure~\ref{fig:design} and executes a continuous analysis loop to recommend ideal number of GPUs required for multiple variants:
\begin{enumerate}
    \item Metrics Collection: It aggregates per-replica metrics from all pods, filtering specifically for the inference engine's custom metric endpoints.
    \item Saturation Thresholds: Scaling is triggered by specific saturation signals defined in the scaling policy. If queue depth or KV cache usage exceeds the defined threshold, the system identifies a need for efficient scale-up.
    \item Safety Nets: The engine calculates a safety net target to prevent scale-to-zero in case of metric pipeline failure, guarding system availability.
\end{enumerate}

\subsection{Global Optimizer}
The Optimization Engine operates in two distinct modes, configurable via an environment variable specified at boot time, directly addressing the community requirements for versatile deployment environments:
\begin{enumerate}
    \item Unconstrained Mode (Default): The optimization problem is treated as separable. The engine solves for the optimal allocation for each deployment independently, assuming infinite lower-layer capacity. This mode targets cloud-native environments backed by auto-scaling node pools, optimizing purely for SLO adherence and cost efficiency.
    \item Constrained Mode: The engine respects global cluster resource constraints (e.g., a fixed 64-GPU cluster). This is the primary operating mode where the Global Optimizer acts as an arbiter. In this mode, multiple variants concurrently request optimal resources based on their individual traffic patterns. However, if summing these requests exceeds cluster capacity, the global optimizer utilizes a \textit{Greedy-by-Saturation} solver to allocate scarce resources. The algorithm filters variants requesting a scale-up and sorts them primarily by their remaining spare capacity ($\delta$) in ascending order, ensuring deployments closest to physical saturation (and nearest to SLO violation) are prioritized for immediate relief. As a secondary tie-breaker, the solver sorts by \texttt{variantCost} in ascending order. GPUs are iteratively allocated to this prioritized queue until the finite cluster budget is exhausted, intrinsically maximizing both SLO safety and energy efficiency.
\end{enumerate}

\subsection{Readiness Gates \& Target Resolution}
To avoid thrashing (scaling up and down rapidly due to incomplete data), WVA implements strict readiness gates:
\begin{description}
    \item[TargetResolved Condition] \hfill \\ The controller verifies the target Deployment exists before attempting any scaling. If the deployment is missing, the VA enters a \texttt{TargetNotFound} state, pausing all Actuation.
    \item[MetricsAvailable Gate] \hfill \\ The Reconciler validates the \texttt{MetricsAvailable} signal within the \texttt{DecisionCache}. If the Engine cannot reach the metrics backend, the Reconciler refuses to change the replica count, preserving the last known good state. This is crucial for inference workload stability during network partition events.
\end{description}

\subsection{The Reconciler}
Operating as a fundamental Kubernetes control loop primitive, the Reconciler executes periodically to align the cluster's actual state with its desired configuration. It is designed to be extremely lightweight, watching a Go channel \texttt{DecisionTrigger} for optimization events, as well as tracking \texttt{VariantAutoscaling} and \texttt{Deployment} events via event filters (to handle creating/deleting targets and multi-controller isolation). On receiving an event, it peeks at the \texttt{DecisionCache} and updates the \texttt{Status} of the \texttt{VariantAutoscaling} resource with the calculated \texttt{DesiredOptimizedAlloc}. It deliberately \textit{avoids} writing ephemeral metric data to the CRD \texttt{Status}, reducing API server pressure (etcd writes) significantly compared to standard HPA~\cite{hpa} which updates status on every metric poll. This underscores WVA's role as an analytic component. WVA calculates the \textit{desired} state based on complex logic, while relying on the HPA control plane to \textit{actuate} it.

\section{Evaluation}

\subsection{Experimental Setup}
Our experimental methodology employs a two-pronged approach encompassing both deterministic simulation and physical cluster validation. We first utilize the \texttt{llm-d-inference-sim}~\cite{llmd_sim} environment for \mbox{deterministic} control plane verification. The primary goal in these initial experiments is to isolate and validate WVA's \textit{decision logic} independently of hardware variance. Following this, we transition to a physical GPU testbed (Section V.D) to validate these efficiencies under real-world system overheads.

For the initial simulation phase, we run WVA alongside the \texttt{llmd} framework in a KinD~\cite{kind_k8s} environment. We explicitly utilize a simulated environment for this baseline evaluation to rigorously isolate WVA's control-loop decision latency from the uncontrollable performance variance of physical hardware. To replicate production metric emission patterns and queue behaviors, we use the aforementioned lightweight simulator, \texttt{llm-d-inference-sim}~\cite{llmd_sim}. We deploy the Qwen/Qwen3-0.6B model, utilizing \texttt{guidellm}~\cite{guidellm} as our commercial-grade load generator. Across all experiments, both simulated and physical, tests are run for 10 minutes (600 seconds) to observe sustained steady-state stability.

\begin{itemize}
    \item Cluster: 3 worker nodes (simulating a mix of A100 and H100 nodes).
    \item Workload: Synthetic load generator sending HTTP requests to trigger queue buildup.
\end{itemize}

To simulate realistic production variability across all experiments, \texttt{guidellm} requests follow a bounded normal distribution for token lengths:
\begin{itemize}
    \item Input Tokens: $\text{min}=10, \text{max}=8192, \text{mean}=4096, \text{stdev}=2048$.
    \item Output Tokens: $\text{min}=10, \text{max}=2048, \text{mean}=1024, \text{stdev}=512$.
\end{itemize}

Autoscaling parameters were configured as follows:
\begin{itemize}
    \item WVA: KV Cache Threshold $\tau_{kv} = 0.8$ ($80\%$), Queue Length Threshold $\tau_{q} = 5$, KV Spare Trigger $\gamma_{kv} = 0.3$ ($30\%$).
    \item HPA: Target Avg. Queue Depth = 3, Target Avg. KV Usage = 0.5 ($50\%$).
    \item \texttt{llmd} scheduler (EPP - endpoint picker): scorers are used with default weights (\texttt{queue-scorer} 1, \texttt{kv-cache-utilization-scorer} 1, and \texttt{prefix-cache-scorer} 1).
\end{itemize}
To rigorously simulate hardware memory constraints underlying the reactivity tests, the inference engine was configured with a physically constrained KV cache (\texttt{--enable-kvcache}, \texttt{--kv-cache-size=1024}, \texttt{--block-size=16}) yielding $16,384$ total tokens. This configuration ensures a rigorous stress test of both saturation-based (WVA) and utilization-based (HPA) scaling strategies under high-variance conditions, alongside realistic traffic distribution mediated by the \texttt{llmd} scheduler (EPP - endpoint picker).
\noindent We configured two \texttt{VariantAutoscaling} resources sharing a single logical model identifier (\texttt{llama3-70b}):
\begin{itemize}
    \item va-a100: Targets the \texttt{deployment\allowbreak-a100} (simulated A100s) with a Cost of 1.0.
    \item va-h100: Targets the \texttt{deployment\allowbreak-h100} (simulated H100s) with a Cost of 2.5.
\end{itemize}
This configuration dictates that the optimizer should prioritize the cheaper \texttt{va-a100} variant until saturated before spilling over to \texttt{va-h100}.

\subsection{Experiment 1: Saturation Reactivity}
\begin{figure}[!htbp]
  \centering
  \includegraphics[width=0.85\linewidth]{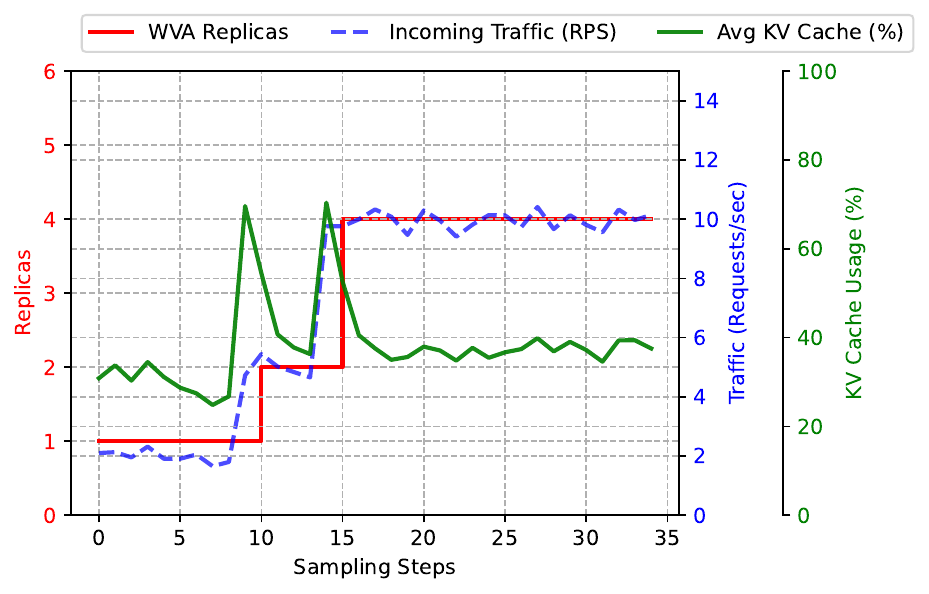}
  \caption{Saturation Reactivity: WVA response time to an unpredicted, near-instantaneous onset of high-concurrency requests.}
  \label{fig:reactivity}
\end{figure}

To evaluate WVA's response to an unpredicted, near-instantaneous onset of high-concurrency requests, we inject a synthetic staircase traffic pattern. As illustrated in Figure~\ref{fig:reactivity} (secondary axis), incoming traffic spikes abruptly from a baseline of ${\sim}2$ requests per second (RPS) to ${\sim}5 \text{ RPS}$ at step 9, and again to ${\sim}10 \text{ RPS}$ at step 14. During these spikes, the active replica's KV Cache (tertiary axis) fills rapidly. With an average request footprint of ${\sim}5,120$ tokens (derived from the mean $4,096$ input and $1,024$ output tokens), the simulated inference server can physically process a maximum of 3 concurrent requests ($3 \times 5,120 = 15,360$ tokens, or $93\%$ utilization) before exhausting memory. By retaining the default \texttt{--max-num-seqs=256}, we ensure the software batch limit does not mask this physical bottleneck, mathematically guaranteeing that WVA's $80\%$ KV-cache threshold ($\tau_{kv} = 0.8$) is cleanly breached by the 3rd concurrent request. As soon as WVA provisions a new replica in response to this loss of headroom, the \texttt{llmd} scheduler (EPP - endpoint picker) instantly routes new load to the fresh capacity, causing the average KV cache utilization to drop back to safe levels.

The primary value proposition demonstrated here is the proactive efficacy of WVA's \textit{Headroom-Based Scaling}. Traditional configurations scaling directly on engine metrics often target an average threshold (e.g., attempting to keep average queue depth at 3). This reactive approach intrinsically necessitates that the target is breached before a deficit is registered, meaning latency degradation has already occurred. Conversely, Figure~\ref{fig:reactivity} shows WVA's replica count (primary axis) stepping up synchronously with the traffic onset to preempt critical system saturation. Because WVA calculates a deterministic safety headroom ($\delta$) based on the physical limits of the KV cache rather than chasing a moving average of requests, it computes the exact number of required replicas \textit{before} the transient spike can exhaust the engine's internal queues. This guarantees that sufficient compute capacity is provisioned to proactively intercept the burst and protect tail-latency SLOs.

\subsection{Experiment 2: Cost-Aware Scaling} \label{sec:experiment-cost}
\begin{figure}[!htbp]
  \centering
  \includegraphics[width=0.85\linewidth]{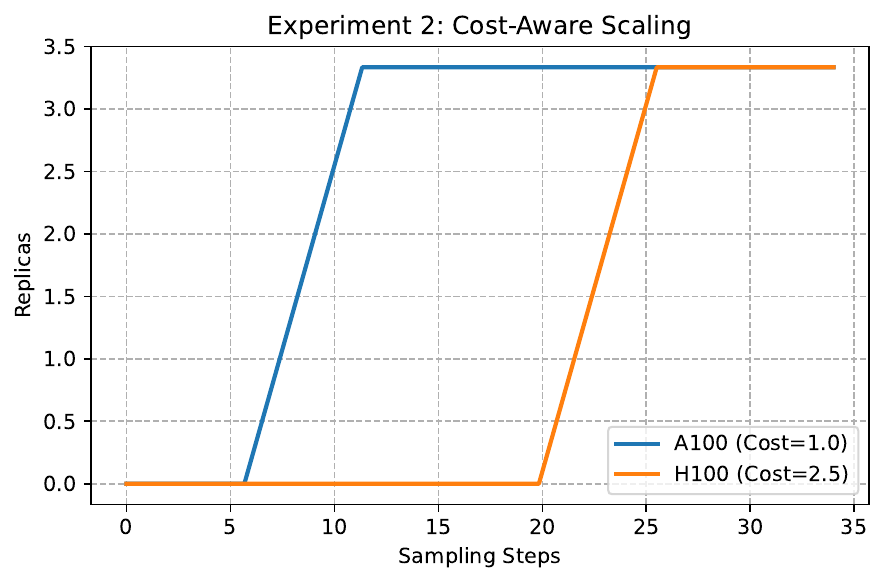}
  \caption{Cost-Aware Scaling: A100s scale up before H100s to minimize inference cost under ramp-up load.}
  \label{fig:cost_scaling}
\end{figure}

We deploy two variants: \texttt{A100} (Cost=1.0) and \texttt{H100} (Cost=2.5). Under a ramp-up load, we observe the replica counts. Our results show that \texttt{A100} replicas scale up first. \texttt{H100} replicas only scale up once \texttt{A100} capacity is saturated or if the rate of arrival exceeds what the \texttt{A100} pool can absorb. This cost-aware tiering inherently provides energy optimization; while the \texttt{H100} offers superior performance-per-watt when fully saturated \cite{nvidia_h100}, the older-generation \texttt{A100} hardware possesses a significantly lower absolute Thermal Design Power (TDP) footprint (up to 400W vs 700W for the \texttt{H100} SXM \cite{nvidia_a100}). By prioritizing the \texttt{A100} for baseline traffic where maximizing throughput per GPU is less critical, WVA avoids the high idle and partial-load power penalties associated with keeping powerful accelerators active, thereby preventing wasteful homogeneous scaling.

\subsection{Physical Cluster Validation}
Completing our two-pronged evaluation strategy, we transition from the isolated simulation environment to validate WVA's behavior under real-world infrastructure constraints and network overheads. We deployed WVA on a large-scale homogeneous physical OpenShift cluster equipped with 200 NVIDIA H100 GPUs, allotting a controlled 10-node sub-pool for this experiment serving the \texttt{Qwen/Qwen3-0.6B} model (configured with \texttt{--max-num-seqs=16000}). We compared WVA against the previously detailed production-grade HPA configuration utilizing the \texttt{llm-d-benchmark}~\cite{llmd_benchmark} suite specifically executing \texttt{guidellm}. The \texttt{llmd} scheduler's \texttt{EndpointPickerConfig} was tailored for this hardware, utilizing \texttt{queue-scorer} (weight $2$), \texttt{kv-cache-utilization-scorer} (weight $2$), and \texttt{prefix-cache-scorer} (weight $3$). The workload for this physical validation specifically tests the llm-d well-lit path configuration~\cite{llmd_well_lit} by applying constant traffic patterns at stepped rates of $[2, 3, 5, 6]$ requests per second (RPS).
\subsubsection{Throughput Stability}
\begin{figure}[!htbp]
  \centering
  \includegraphics[width=0.85\linewidth]{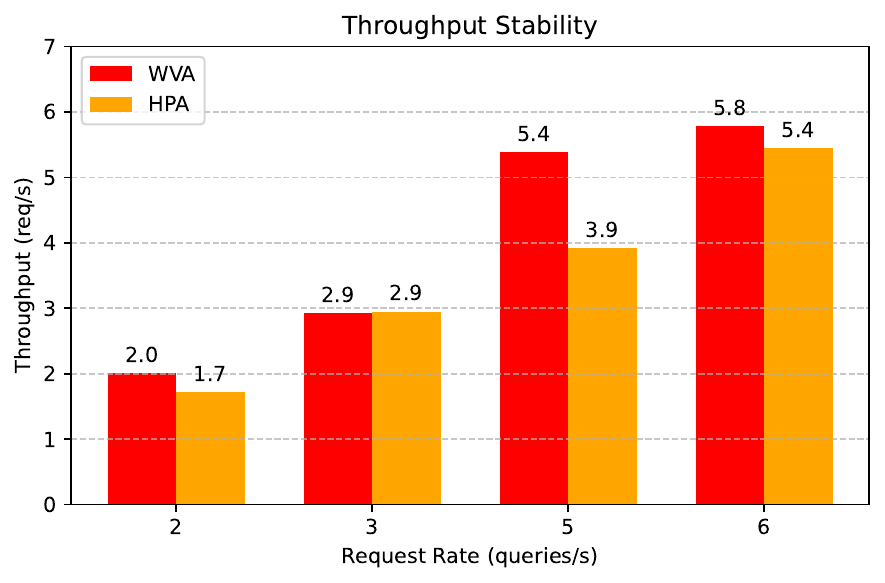}
  \caption{Throughput Stability: WVA (Red) maintains higher, more consistent throughput compared to HPA (Orange).}
  \label{fig:throughput}
\end{figure}

As established in Section V.B, HPA's foundational logic targets an average utilization state, a reactive approach that intrinsically prevents it from scaling in time to intercept abrupt load steps. This introduces unavoidable queuing latency and premature throttling, resulting in degraded overall throughput. In contrast, WVA's \textit{Headroom-Based Scaling} computes the required capacity \textit{proactively}. By maintaining the aforementioned proactive spare capacity buffer ($\delta$) calibrated to absorb these stepped increases, WVA scales up \textit{before} the buffers are exhausted. Consequently, WVA sustains significantly higher effective throughput (peaking at a 37\% improvement at 5 RPS, see Figure~\ref{fig:throughput}) by avoiding the queuing degradation typical of reactive policies.

\subsubsection{Request Stability}
\begin{figure}[!htbp]
  \centering
  \includegraphics[width=0.85\linewidth]{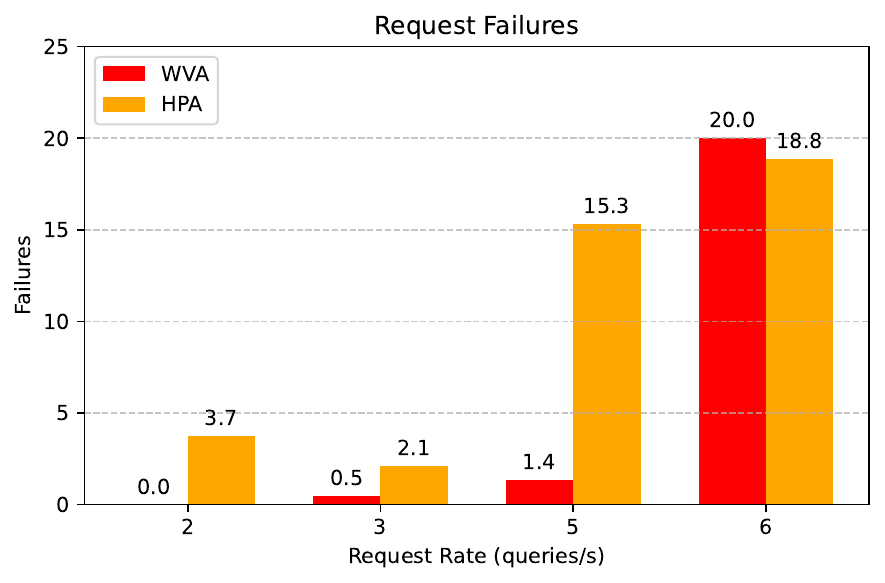}
  \caption{Request Stability: WVA (Red) minimizes drops via saturation buffers and safe termination, while HPA (Orange) incurs failures due to throttling and instability.}
  \label{fig:failures}
\end{figure}

Figure~\ref{fig:failures} highlights the system's robustness under load. HPA exhibits a baseline of request failures even at moderate loads (2--5 RPS), primarily driven by two factors:
\begin{enumerate}
    \item Saturation Rejection: HPA targets average utilization, keeping the system near capacity. Abrupt load steps trigger \texttt{llmd}'s admission control, rejecting requests (HTTP 429/503) due to full KV cache or queues.
    \item Scale-Down Instability: Signal noise can trigger premature scale-down actions by HPA, terminating pods that are still processing active requests.
\end{enumerate}
In contrast, the aforementioned proactive capacity buffer ($\delta$) naturally absorbs these stepped increases, preventing saturation-based rejection. Furthermore, WVA's \textit{Fragmentation-Aware Scale Down} ensures that no pod is terminated until it is strictly drained. This dual approach eliminates failures at sub-saturation loads (maintaining $<1.5$ drops/s at 5 RPS while HPA spikes to $>15$ drops/s).

\subsubsection{Efficiency \& Replica Count}
WVA consistently maintained a lower replica count than HPA at moderate load levels. However, a critical divergence occurs between 5-6 RPS: WVA aggressively scales to the maximum limit (10 replicas) while HPA remains at $\approx$9.3 replicas. This overshoot is a deliberate consequence of WVA's \textit{Headroom-Based Scaling}. 

Since the \texttt{llmd} scheduler restricts request admission based on saturation signals (e.g., KV cache fullness), it actively throttles traffic when the system is at capacity. HPA, by targeting average utilization, implicitly keeps the system in this saturated state, causing \texttt{llmd} to reject excess load (lower throughput) to protect existing requests. In contrast, WVA provisions spare capacity $\delta$, ensuring the system remains slightly \textit{under-saturated}. This allows \texttt{llmd} to admit the full burst of traffic (5.78 req/s vs 5.45 req/s).

This load level represents a critical saturation point where the workload exceeds the physical capacity of the maximum allowable replicas (10). While HPA throttles admission early (resulting in lower throughput) and rarely requests the maximum units, WVA's \textit{Headroom-Based Scaling} correctly calculates that more than 10 replicas are required to maintain the safety buffer $\delta$. Because the orchestrator caps the deployment at 10, WVA successfully scales to the hard limit and maximizes available service. By processing closer to the hardware's physical limit without the buffer, WVA achieves significantly higher throughput but incurs a marginal increase in tail-end failures compared to the more conservative HPA. This demonstrates WVA's design philosophy maximizing useful service (Goodput) even when physical constraints prevent ideal headroom provisioning.

\subsection{Latency Distribution}
\begin{figure}[!htbp]
  \centering
  \includegraphics[width=0.85\linewidth]{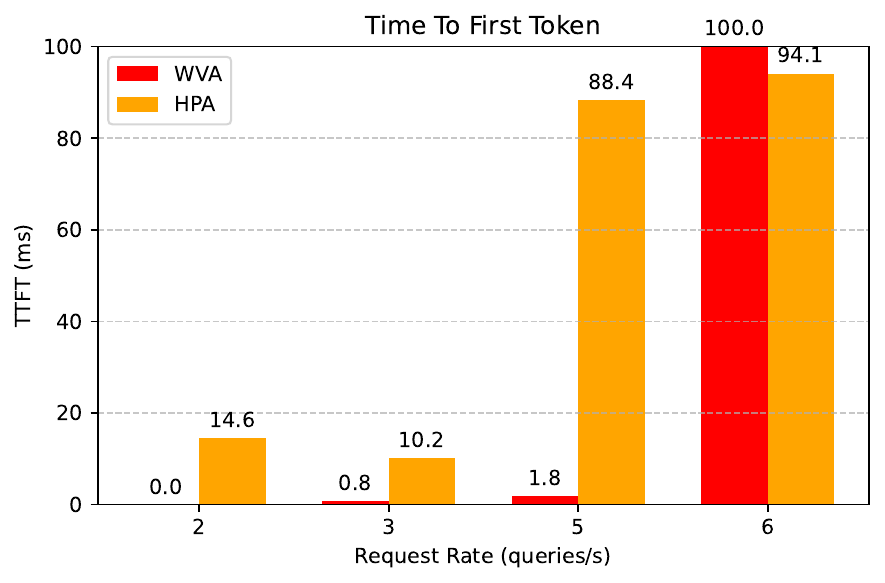}
  \includegraphics[width=0.85\linewidth]{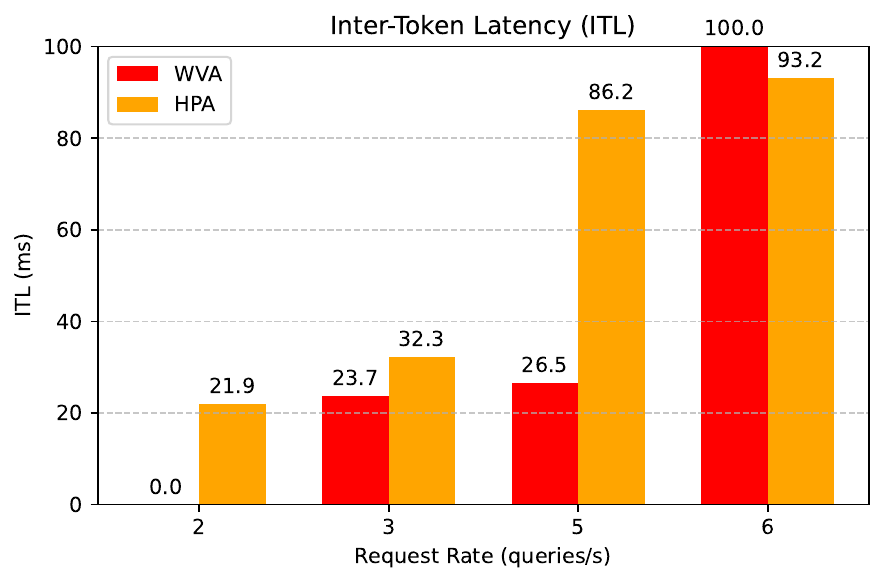}
  \caption{Mean Latency Distribution: TTFT (Top) and ITL (Bottom). WVA (Red) maintains a stable mean latency and protects aggregate SLOs comparable to HPA (Orange) despite driving higher admission rates.}
  \label{fig:latency}
\end{figure}

Latency metrics confirm that WVA's scaling strategy successfully protects the application's Service Level Objectives (SLOs) under moderate, unconstrained load. For interactive LLM serving, the primary SLOs are minimizing Time To First Token (TTFT)---which encompasses both compute time and any queueing delay incurred in the scheduler---and Inter-Token Latency (ITL). Figure~\ref{fig:latency} shows the distribution of mean latencies for both metrics. Both mean TTFT and mean ITL remain stable for WVA (Red), staying well within acceptable SLO boundaries up to 5 RPS. By proactively provisioning the $\delta$ safety buffer, WVA ensures that incoming requests are immediately assigned to compute resources, effectively eliminating the queueing delay that governs mean TTFT spikes in reactive systems.

A slight elevation in overall mean latency and a corresponding increase in the SLO violation risk is observed at 6 RPS; this is an expected trade-off as WVA's \textit{Headroom-Based Scaling} is restricted by the 10-replica hard deployment limit. Because WVA can no longer provision spare capacity, the existing replicas are forced to absorb the excess traffic without the protective $\delta$ buffer, causing higher contention for compute resources. In the HPA configuration, the system stabilizes at a lower replica count ($\approx 9.3$) which frequently triggers \texttt{llmd}'s internal saturation thresholds (e.g., KV cache limits), forcing the scheduler to reject excess requests (throttling). High rejection rates artificially lower the mean latency for admitted requests by reducing concurrent load, whereas WVA maximizes admission at the cost of increased resource contention.

As shown in the physical validation results, WVA demonstrates tangible value over traditional CPU-centric heuristics. By leveraging deep application introspection (queue depth and KV cache usage), WVA achieves significant efficiency gains while maintaining SLOs. Unlike static baselines or generic autoscalers, WVA dynamically navigates the trade-off between latency and cost, proving its ability to solve the multi-dimensional optimization problem central to modern LLM serving.

\section{Related Work}

\subsection{Event-Driven Autoscaling}
KEDA (Kubernetes Event-driven Autoscaling) \cite{keda} extends HPA by allowing scaling based on event sources (typically message queues like Kafka or RabbitMQ). While KEDA improves reactivity over standard HPA by scaling on queue depth (similar to WVA), it remains fundamentally pod-centric. KEDA scales individual ScaledObjects in isolation, lacking a global optimization context. It cannot, for example, decide to scale up a cheaper variant instead of an on-demand one based on global cluster capacity or cost constraints. Furthermore, unlike KEDA which scales on \textit{raw} events, WVA scales on \textit{processed} saturation signals derived from deep application introspection (e.g., KV cache utilization). This ensures that scaling decisions are strictly engine-aware and hardware-efficient, avoiding oscillation from noisy raw metrics.

\subsection{Serverless and Request-Based Scaling}
Knative Pod Autoscaler (KPA) \cite{kpa} introduces concurrency-based scaling, which is highly effective for request-driven serverless workloads. KPA scales based on the number of concurrent requests per pod (concurrency target). However, as established in Section II, a single LLM request can hold significant resources (KV cache) for a long duration. WVA's approach of directly monitoring the inference engine's internal state (KV cache usage, queue saturation) provides a higher-fidelity signal for this specific domain than generic request concurrency.

\subsection{LLM-Centric Scheduling and Scaling}
Recent research has proposed several specialized systems for LLM and multimodal serving; while not an exhaustive survey, the following represent key advancements in the domain. Orca \cite{orca} pioneered iteration-level scheduling to handle variable-length sequences. Chiron \cite{chiron} introduces hierarchical autoscaling that utilizes backpressure signals (queue size, SLO adherence) to optimize resource allocation for large models. HeteroScale \cite{heteroscale} addresses the challenges of disaggregated prefill-decode architectures, coordinating scaling across heterogeneous hardware pools to balance network bottlenecks and computational demands. Similarly, ModServe \cite{modserve} extends disaggregation concepts to multimodal models, offering modality-aware scheduling and scaling.

While these systems excel at micro-level optimization via specialized engine modifications \cite{orca, heteroscale}, WVA approaches the problem from the cluster orchestration layer. As detailed in Section \ref{sec:headroom}, WVA formalizes scaling not as hierarchical backpressure \cite{chiron}, but as a mathematically pure constraint satisfaction problem defining optimal system state ($Capacity_{target} = Load_{current} + \delta_{safety}$) without altering the core scheduling loop. 

Furthermore, unlike systems bounded to specific architectures, WVA introduces the \textit{Variant} as a Kubernetes-native abstraction. As demonstrated in Section \ref{sec:experiment-cost}, this decoupled control plane enables cluster-aware hardware tiering—prioritizing absolute power efficiency (A100) for baseline traffic while reserving high performance-per-watt (H100) for saturated bursts. Ultimately, WVA's novelty lies in bridging deep application introspection with generalized cloud-native orchestration, achieving domain-specific optimization while preserving strict modularity and extensibility.

\section{Future Work}
While WVA successfully addresses the heterogeneity and cost-efficiency of LLM serving through its saturation-based reactive scaling, it remains bounded by the latency of reaction. Scaling actions are triggered only \textit{after} queues begin to build up or KV cache thresholds are crossed. For bursty workloads, this can still result in transient latency spikes.

Future work will focus on moving to better proactive scaling via traffic shape prediction. By integrating time-series forecasting models (such as LSTMs or Transformer-based predictors) into the optimization engine, WVA can learn historical traffic patterns and predict varying arrival rates. This will enable the system to pre-provision capacity ahead of anticipated load spikes, effectively minimizing cold-start latency and ensuring consistent performance during rapid demand shifts.

Another critical avenue for future development is energy-aware optimization. As AI workloads consume an increasing portion of global data center power, optimizing for energy efficiency becomes paramount. We plan to extend the global optimizer to ingest real-time power consumption metrics and carbon intensity signals. This will enable WVA to make scaling decisions that prioritize not only cost and latency but also environmental impact, for example by provisioning capacity for non-critical batch workloads in regions with greener energy or preferentially utilizing hardware variants with better performance-per-watt profiles.

Finally, we aim to address the distinct resource requirements of inference phases, further improving independent scaling of prefill and decode workers. Additionally, we plan to evaluate WVA in conjunction with the emerging flow control capabilities within the \texttt{llmd} scheduler (EPP) to provide even tighter tail-latency guarantees under extreme load.

\section{Conclusion}
In this work, we presented WVA, a system that introduces the concept of \textit{variants} to optimize AI inference at scale. By dynamically orchestrating these variants and maintaining proactive safety headroom, WVA effectively avoids the queuing degradation typical of reactive systems, thereby sustaining higher effective throughput and protecting tail-latency SLOs compared to traditional scaling approaches. While WVA is implemented for \texttt{llmd}, the pattern of Deep Vertical Integration---where the autoscaler and the engine share a common language of saturation---is a generalizable lesson for all future AI serving systems. We envision WVA as a foundation for sustainable AI infrastructure. WVA is in active development as a core capability of \texttt{llmd}; we invite the open-source and research community to collaborate on future extensions by engaging with the project on GitHub at \url{https://github.com/llm-d/llm-d-workload-variant-autoscaler}.

\section*{Acknowledgments}
We sincerely thank Kaushik Mitra (Google) for his valuable insights and feedback during the early design phase of WVA. We are also grateful to Andy Anderson (IBM Research) for his crucial support in establishing the CI pipelines on our GPU clusters.

\bibliographystyle{IEEEtran}
\bibliography{references}

\end{document}